\documentclass[english,numberedappendix]{emulateapj}
\usepackage[T1]{fontenc}
\usepackage{babel}
\usepackage{array}
\usepackage{amssymb}
\usepackage{graphicx}
\usepackage[unicode=true]
 {hyperref}
\usepackage{breakurl}

\makeatletter

\providecommand{\tabularnewline}{\\}

\usepackage{amsmath}
\slugcomment{}
\shorttitle{Photospheric radius expansion in superburst precursors}
\shortauthors{Keek}

\makeatother

\begin{document}

\title{Photospheric radius expansion in superburst precursors from neutron
stars}

\author{L.~Keek}

\affil{National Superconducting Cyclotron Laboratory, Department of Physics
\& Astronomy, and Joint Institute for Nuclear Astrophysics, Michigan
State University, East Lansing, MI 48824, USA}

\email{keek@nscl.msu.edu}
\begin{abstract}
Thermonuclear runaway burning of carbon is in rare cases observed
from accreting neutron stars as day-long X-ray flares called superbursts.
In the few cases where the onset is observed, superbursts exhibit
a short precursor burst at the start. In each instance, however, the
data was of insufficient quality for spectral analysis of the precursor.
Using data from the propane anti-coincidence detector of the PCA instrument
on \emph{RXTE}, we perform the first detailed time resolved spectroscopy
of precursors. For a superburst from 4U~1820--30 we demonstrate the
presence of photospheric radius expansion. We find the precursor to
be $1.4-2$ times more energetic than other short bursts from this
source, indicating that the burning of accreted helium is insufficient
to explain the full precursor. Shock heating would be able to account
for the lacking energy. We argue that this precursor is a strong indication
that the superburst starts as a detonation, and that a shock induces
the precursor. Furthermore, we employ our technique to study the superexpansion
phase of the same superburst in greater detail.
\end{abstract}


\section{Introduction}

\label{sec:Introduction}Superbursts are rare day-long X-ray flares
observed from accreting neutron stars in low-mass X-ray binaries (LMXBs,
\citealt{Cornelisse2000,Strohmayer2002}). They are attributed to
runaway thermonuclear fusion of carbon in an approximately $100\,\mathrm{m}$
thick layer below the neutron star surface (\citealt{Cumming2001}).
Superbursts share some properties of the frequently detected short
Type I X-ray bursts (durations of $\sim10-100\,\mathrm{s}$), which
result from runaway burning of hydrogen and/or helium accreted onto
the neutron star from the binary companion (e.g., \citealt{Lewin1993}).
For example, both have a fast rising light curve, followed by a slow
decay during which the X-ray spectrum exhibits cooling. Superbursts,
however, are $1000$ times more energetic, last $1000$ times longer,
and occur $1000$ times less frequently than the short bursts (e.g.,
\citealt{Keek2008int..work}).

Currently $22$ (candidate) superbursts are known from $13$ sources
(see, e.g., \citealt{Keek2012} for an overview). All detections are
performed with observatories in low earth orbit, which implies observations
are frequently interrupted by Earth occultations and passages through
the South-Atlantic Anomaly. Often the start of a superburst is not
observed. In some cases this makes it difficult to prove the superburst
nature of an event: these are referred to as candidates (e.g., \citealt{Zand2004}).
Only in $6$ cases is there an unambiguous detection of the onset.
In all these cases a precursor burst is present at the superburst
start (\citealt{Strohmayer2002,Strohmayer2002a,Zand2003,Zand2004}).
In previous studies the term ``precursor'' has also been used to
describe a burst that precedes the superburst by two minutes (\citealt{Kuulkers2002ks1731}),
or when two of these events are visible in the superburst rise (\citealt{Strohmayer2002}).
Because these likely have a different physical origin, we restrict
our study to those precursor bursts that immediately precede and transition
into the superburst.

The most detailed precursor observations have been performed with
the Proportional Counter Array (PCA) on the \emph{Rossi X-ray Timing
Explorer} (\emph{RXTE}), and are from superbursts of 4U~1820--30
(\citealt{Strohmayer2002}) and 4U~1636--53 (\citealt{Strohmayer2002a}). 

\object{4U~1636--53} is a prolific burster that has exhibited many
different modes of thermonuclear burning of hydrogen and helium (e.g.,
\citealt{Hoffman1977,Paradijs1986,Revnivtsev2001}), as well as carbon
burning in 4 superbursts (\citealt{Wijnands2001sb,Strohmayer2002a,2004Kuulkers,Kuulkers2009ATel}). 

\object{4U~1820--30} is an ultracompact X-ray binary (UCXB, \citealt{1987Stella}),
implying that the material accreted from the donor star is likely
helium rich (\citealt{Rappaport1987}, see also \citealt{2003Cumming}).
Its persistent flux is observed to vary on a timescale of approximately
$171$~days (\citealt{Priedhorsky1984,cho1}), and X-ray bursts are
only detected during periods of low flux (e.g., \citealt{Cornelisse2003}).
Two (candidate) superbursts have been observed (\citealt{Strohmayer2002,Zand2011ATel}).
The first superburst from this source was observed with the PCA, and
used to study the changing properties of the accretion disk during
the superburst (\citealt{Ballantyne2004}), the occurrence of superexpansion
of the photosphere (\citealt{Zand2010}), and achromatic variability
in the light curve (\citealt{Zand2011}).

The precursors resemble short X-ray bursts, and the light curve exhibits
two peaks, similar to photospheric radius expansion (PRE) bursts (\citealt{Tawara1984,Lewin1984}).
During powerful X-ray bursts, when the luminosity reaches the Eddington
limit, the photon pressure is sufficient to (temporarily) push out
the photosphere, causing the surface temperature to drop. The lower
temperature results in a smaller part of the black-body flux to fall
within the energy band in which many X-ray detectors are sensitive,
producing a dip in the light curve. Unfortunately no spectral data
of sufficient quality have been available to confirm the PRE nature
of superburst precursors.

An alternative explanation for the double peaked nature of precursors
has been put forth. Carbon burning in a superburst that proceeds as
a detonation drives a shock to the surface (\citealt{Weinberg2006sb}),
where it produces a shock-breakout peak in the light curve, followed
by a burst from the shock-ignited burning of hydrogen and helium in
the atmosphere. \citet{Weinberg2007} argue that the two peaks are
separated by a thermal timescale of seconds, similar to the two peaks
observed in precursors. \citet{Keek2011}, however, show that the
fallback of the shock-accelerated atmosphere alone dissipates enough
energy to power a bright precursor on a dynamical timescale of $10\,\mu\mathrm{s}$.
At the same time any hydrogen or helium present in the atmosphere
ignites and contributes to the precursor energetics (\citealt{Keek2012}).
In this scenario the shock breakout and the subsequent burst are too
close in time to be distinguished as two peaks, but the burst is bright
enough to cause radius expansion.

To test the different models for superburst precursors, we devise
a technique using anti-coincidence data of the PCA to perform time
resolved spectroscopy on the two superbursts observed with this instrument
from, respectively, 4U~1820--30 and 4U~1636--53. This is the first
time precursors are studied in detail, and we discuss the implications
for the nature of precursors and of superburst ignition.

\section{Observations}

\emph{RXTE} (\citealt{Levine1996}) was launched in December 1995
and observed the X-ray sky until January 2012. It carries the PCA
(\citealt{Jahoda2006}), which has a band pass of 2 to 60~keV, and
a total collecting area of $6500\,\mathrm{cm^{2}}$. As part of its
science program the PCA frequently observed LMXBs, and it detected
thousands of short X-ray bursts (e.g., \citealt{Galloway2008catalog})
as well as the two mentioned superbursts.

The PCA consists of 5 identical collimated Proportional Counter Units
(PCUs). Each PCU contains three xenon-filled layers, which act as
the main counter, and are sensitive in the 2--60~keV energy range.
On top of the xenon detector is a propane-filled anti-coincidence
detector. It provides additional sensitivity between 1.8--3.5~keV
with peak sensitivity at $2.5\,\mathrm{keV}$%
\footnote{We refer to \href{http://heasarc.gsfc.nasa.gov/docs/xte/appendix_f.html}{http://heasarc.gsfc.nasa.gov/docs/xte/appendix\_{}f.html}
for XTE Technical Appendix F.%
}, but little calibration information is available. Data from the xenon
layers are typically available per layer and PCU at high time resolution
and with spectral information, whereas for the propane layer only
a count rate for all PCUs combined is available at $0.125\,\mathrm{s}$
time resolution without spectral information.

At the time of both superburst observations there were problems that
resulted in a loss of data. At the time of the 4U~1820--30 superburst
observation, \emph{RXTE} was troubled by a malfunction that prevented
part of the data from being transmitted to the ground (\citealt{Strohmayer2002}).
During the 4U~1636--53 superburst the data buffer overflowed, causing
a loss of data (\citealt{Strohmayer2002a}). Most notably, high time
resolution spectral data were lost in both cases. During the superburst
rises only Standard1 and Standard2 data are available, which have
a time resolution of $0.125\,\mathrm{s}$ and $16\,\mathrm{s}$, respectively.
Only the latter mode contains spectral information, but its time resolution
exceeds the duration of the precursors. We are, therefore, limited
to using Standard1 data. Additionally, the propane count rate is available
at $0.125\,\mathrm{s}$ time resolution.

Because of the different energy responses of the xenon and propane
layers, count rates from the two can be regarded as representing two
energy channels. This allows for fitting a spectral model with two
parameters, such as the black-body model that is often used to describe
burst spectra (\citealt{swank1977}). The propane signal, however,
is not calibrated for this purpose (\citealt{Jahoda2006}). To investigate
the energy response in these two ``channels'' to a black-body spectrum,
we perform time resolved spectroscopy on several short bursts from
the two superbursters for which event data with a minimum time resolution
of $125\,\mathrm{\mu s}$ and with $64$ energy channels are available. 

We select short bursts from version 0.51 of the Multi-INstrument Burst
Archive%
\footnote{See \href{http://users.monash.edu.au/~dgallow/minbar}{http://users.monash.edu.au/$\sim$dgallow/minbar}
for more details.%
} (MINBAR, \citealt{Keek2010}), which contains the results of the
analysis of 4,192 bursts observed from 72 sources with the PCA (\citealt{Galloway2008catalog})
and the Wide-Field Camera's on the \emph{BeppoSAX} observatory (\citealt{Cornelisse2003}).
For the purpose of this study we only use PCA bursts.

The gain settings of the PCUs were changed several times during the
mission. To be able to compare count rates, we restrict ourselves
to bursts that were observed with the same detector gain as the respective
superbursts. Furthermore, the gain was slightly different for each
PCU (up to $1\%$, \citealt{Jahoda2006}). The propane data, however,
are stored as an aggregate, and not per PCU. We, therefore, use the
propane and xenon data of all active PCUs.

The number and configuration of active PCUs changes per observation,
and due to the impact of micrometeorites PCU 0 and 1 lost their propane
layer on, respectively, May 13 2000 and December 25 2006 (\citealt{Jahoda2006}).
We, therefore, report xenon and propane count rates per active detector.
An active detector that has lost its propane layer is still counted
as an active detector for determining the xenon count rate, but not
for the propane rate. 

There are small changes in the gain and instrument response over time.
For example, xenon is known to have leaked into the propane layer.
When analyzing the short bursts, we generate response matrices for
each burst to take into account the instrument response and the configuration
of active PCUs at a particular time.

For spectra and count rates we subtract the background measured during
a time interval before each (super)burst. This yields the net burst
rates and spectra, and removes the astrophysical background as well
as the instrumental background. Here we assume both backgrounds remain
constant during the burst. We further discuss this in Section~\ref{sub:method improvements},
but especially in the burst peak the signal is expected to be dominated
by the burst, so small changes in the background should not be detrimental
to our analysis.

All extracted data products, including spectra and light curves, are
corrected for dead time using the prescription provided by the instrument
team. Data required for this correction are available at a resolution
of $0.125\,\mathrm{s}$. When calculating the correction we take into
account that some of the active PCUs may have lost their propane layer.

\section{Results}

\subsection{4U~1820--30}

\begin{table}
\caption{\label{tab:bursts_1820}13 PRE Bursts from 4U~1820--30 and a Superburst
Precursor (Bottom)}
\begin{tabular}{r@{\extracolsep{0pt}.}lr@{\extracolsep{0pt}.}l>{\centering}p{1.5cm}>{\centering}p{2.2cm}}
\hline 
Obs& ID & \multicolumn{2}{c}{Time} & Fluence  & Xenon counts \tabularnewline
\multicolumn{2}{c}{} & \multicolumn{2}{c}{(MJD)} & ($10^{-7}$ erg cm$^{-2}$) & ($10^{4}$ c cm$^{-2}$ PCU$^{-1}$)\tabularnewline
\hline 
\multicolumn{2}{c}{40017-01-24-00} & 52794&73813 & $3.6\pm0.1$ & $3.20\pm0.08$\tabularnewline
\multicolumn{2}{c}{70030-03-04-01} & 52802&07557 & $4.0\pm0.1$ & $3.37\pm0.07$\tabularnewline
\multicolumn{2}{c}{70030-03-05-01} & 52805&89566 & $4.1\pm0.2$ & $3.31\pm0.08$\tabularnewline
\multicolumn{2}{c}{90027-01-03-05} & 53277&43856 & $3.7\pm0.1$ & $3.24\pm0.08$\tabularnewline
\multicolumn{2}{c}{94090-01-01-02} & 54948&82124 & $3.2\pm0.2$ & $3.01\pm0.14$\tabularnewline
\multicolumn{2}{c}{94090-01-01-05} & 54950&70281 & $3.5\pm0.1$ & $3.22\pm0.10$\tabularnewline
\multicolumn{2}{c}{94090-01-02-03} & 54956&77470 & $3.3\pm0.1$ & $3.08\pm0.11$\tabularnewline
\multicolumn{2}{c}{94090-01-02-02} & 54958&73998 & $3.2\pm0.2$ & $2.98\pm0.14$\tabularnewline
\multicolumn{2}{c}{94090-01-04-00} & 54978&32149 & $3.7\pm0.2$ & $3.24\pm0.14$\tabularnewline
\multicolumn{2}{c}{94090-01-04-01} & 54978&49489 & $3.7\pm0.2$ & $3.3\pm0.2$\tabularnewline
\multicolumn{2}{c}{94090-01-05-00} & 54981&18728 & $3.8\pm0.1$ & $3.21\pm0.09$\tabularnewline
\multicolumn{2}{c}{94090-02-01-00} & 54994&53418 & $3.2\pm0.2$ & $2.95\pm0.13$\tabularnewline
\multicolumn{2}{c}{94090-02-01-00} & 54994&61301 & $3.6\pm0.1$ & $3.15\pm0.10$\tabularnewline
\hline 
\multicolumn{2}{c}{30057-01-04-08} & 51430&07423 & $6-8$ & $4.59\pm0.06$\tabularnewline
\hline 
\end{tabular}
\end{table}
On 9/9/1999 the PCA detected a superburst from 4U~1820--30 (\citealt{Strohmayer2002}).
The superburst was observed for three hours, and the flux continued
to decrease when the observation ended. Two events with PRE-like peaks
and dips are visible in the light curve. We consider the first to
be the precursor. MINBAR lists $14$ bursts from this source detected
with the PCA, 13 of which were observed with the same high voltage
settings as the superburst (Table~\ref{tab:bursts_1820}). All bursts
exhibit PRE. We perform time-resolved spectroscopy on the bursts to
calibrate the xenon and propane count rates such that these can be
used to analyze the superburst precursor.

\subsubsection{Analysis of short bursts}

Net burst spectra are extracted at a resolution of $0.125\,\mathrm{s}$
in the rise and at the peak of the burst. After the peak we decrease
the time resolution by a factor $2$ every time the count rate per
PCU decreases by $\sqrt{2}$, which ensures similar statistics for
all spectra. Each spectrum is fit with a black body in the $2.5-25\,\mathrm{keV}$
energy range. Absorption by interstellar hydrogen is taken into account
using the cross sections of \citet{1983Morrison}, while fixing the
hydrogen column to $N_{\mathrm{H}}=0.25\times10^{22}$ (\citealt{Asai2000}).
The fit yields the black-body temperature, $kT$, and normalization,
$K_{\mathrm{bb}}\equiv\left(\frac{R}{\mathrm{km}}\right)^{2}\left(\frac{d}{10\,\mathrm{kpc}}\right)^{-2}$,
with $R$ the black-body radius and $d$ the distance to the source.

\begin{figure}
\includegraphics{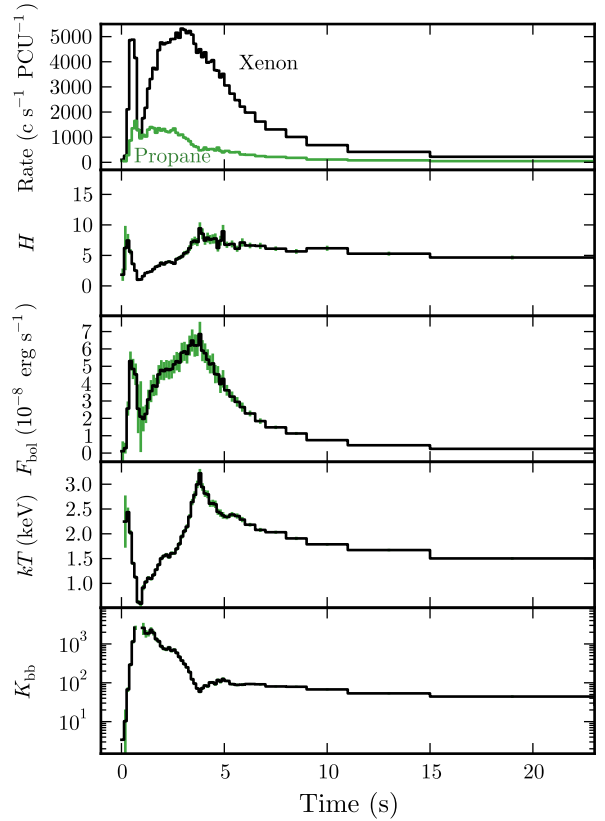}

\caption{\label{fig:PRE-burst} For the X-ray burst on MJD~52794 from 4U~1820--30
the count rate in the xenon and the propane layers, hardness $H$
(ratio of xenon and propane counts), and the results of time resolved
spectroscopy: the unabsorbed bolometric flux $F_{\mathrm{bol}}$,
the black-body temperature $kT$, and the black-body normalization
$K_{\mathrm{bb}}$. For the latter we omit two data points with anomalously
large values and errors. Vertical (green) lines indicate the $1\sigma$
uncertainties per time bin (no uncertainties are indicated for the
rates).}
\end{figure}

For each time bin we extract the count rate in the propane layer and
in the xenon layer (per PCU). We determine the sum of the xenon and
propane count rate, $I$, and the hardness ratio of the two rates,
$H$. An example of the evolution during a burst of these quantities
and the spectral fit parameters is shown in Figure~\ref{fig:PRE-burst}.

4U~1820--30 experiences rather strong radius expansion during its
bursts, which pushes a large part of the emission out of the PCA band.
The peak of the count distribution for a black body lies at an energy
of $E_{\mathrm{max}}\simeq1.59\, kT$. Because of absorption at low
energies, the peak of the spectrum is shifted to slightly higher energies.
When the black body temperature drops below $kT\simeq1.6\,\mathrm{keV}$
the peak is outside the considered band-pass. The fits become worse
around $kT\simeq1.0\,\mathrm{keV}$, resulting in dips and peaks in
the bolometric flux. Therefore, our fits are not reliable for $kT\lesssim1.0\,\mathrm{keV}$.
In further analysis, we use $kT>1.0\,\mathrm{keV}$ when deriving
relations between the different parameters. Furthermore, in the tail
of the bursts the net burst counts are strongly reduced, and the relative
uncertainty in $H$ and $I$ become large. In our analysis we omit
data points where the relative uncertainty in $H$ exceeds $50\%$.

\begin{figure}
\includegraphics{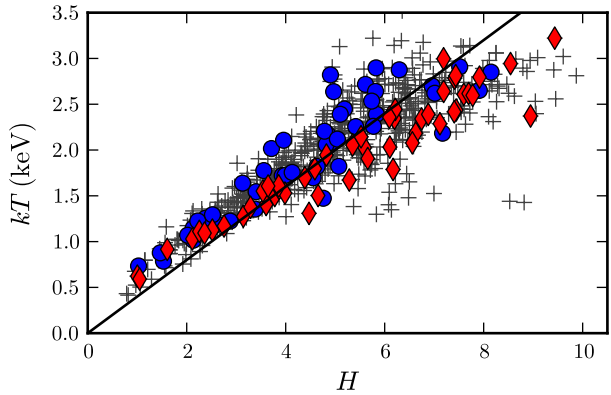}

\caption{\label{fig:Hardness-kT-1}Hardness (ratio of xenon and propane counts)
versus the black-body temperature $kT$ from time-resolved spectroscopy
of 13 PRE bursts. Crosses indicate the data points for all bursts
(no uncertainties implied), whereas disks and diamonds are for two
bursts on MJD~52794.7 and MJD~54994.5, respectively. The solid line
is the best linear fit to the data with $kT>1.0\,\mathrm{keV}$.}
\end{figure}

The hardness ratio traces changes in the spectral shape, which for
a black body is set by $kT$. We determine the relation between $kT$
and $H$ by fitting a straight line (Figure~\ref{fig:Hardness-kT-1}).
To determine the goodness of fit we use the error in $H$, because
it is typically larger than the error in $kT$. To ensure the latter,
we exclude data points where the relative uncertainty in $kT$ exceeds
$10\%$. The best fit line is 
\begin{equation}
kT=(0.40\pm0.06)H\,\mathrm{keV}.\label{eq:hardness_kT-1820}
\end{equation}
To obtain $\chi_{\nu}^{2}=1$ a systematic error of $16\%$ needs
to be added in quadrature to the uncertainty of the data points. This
describes the spread in the distribution of both individual bursts
and all bursts combined. Some individual bursts have a smaller spread
(Figure~\ref{fig:Hardness-kT-1}), but the spread between such bursts
is substantial, so this does not aid us in improving the fit. Furthermore,
the fitted line is defined to go through zero. Including an offset,
the best fit conforms better to the points with $kT\lesssim1.0\,\mathrm{keV}$
(which are excluded from the fit), but the fit statistics are not
improved; the best fit value is within $1\sigma$ consistent with
$0.0$. Furthermore, competition between the fit parameters leads
to large uncertainties in the parameters.

\begin{figure}
\includegraphics{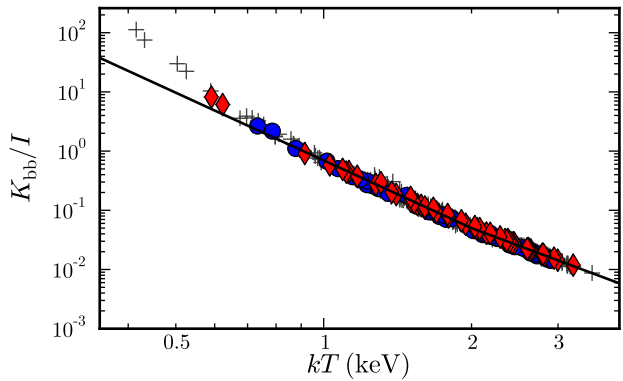}

\caption{\label{fig:kT-ratio-1}Black-body temperature $kT$ versus the ratio
of the black-body normalization $K_{\mathrm{bb}}$ and the total count
rate $I$ for 13 PRE bursts. See Figure~\ref{fig:Hardness-kT-1}
for an explanation of the symbols. The solid line is the best fit
broken power-law to the data with $kT>1.0\,\mathrm{keV}$.}
\end{figure}

Given $kT$, which sets the spectral shape, $K_{\mathrm{bb}}$ is
linearly proportional to $I$. We determine the ratio $K_{\mathrm{bb}}/I$,
and we find that as a function of $kT$ it follows a relation that
is close to a power law (Fig.~\ref{fig:kT-ratio-1}). 
 We fit a broken power law with the break at $2\,\mathrm{keV}$, which
is approximately the center of the region over which we fit, and we
define the power-law normalization at the location of the break. The
best fit is 
\begin{equation}
\frac{K_{\mathrm{bb}}}{I}=(5.0\pm0.4)\times10^{-2}\left(\frac{kT}{2\,\mathrm{keV}}\right)^{-\Gamma},\label{eq:ratio_kT-1820}
\end{equation}
with $\Gamma=3.8\pm0.2$ for $kT<2\,\mathrm{keV}$ and $\Gamma=3.10\pm0.07$
for $kT\ge2\,\mathrm{keV}$. The power-law index is $1.4\,\sigma$
from $3.0$ for $kT>2\,\mathrm{keV}$ (where the peak of the black-body
counts spectrum is within the xenon detector band-pass), which is
the expected value for a Planck counts spectrum. Performing a fit
of a single power law with the index fixed to this value leads, however,
to a larger error in the prefactor. 

For all bursts we determine the total number of xenon counts per PCU
and the unabsorbed bolometric fluence (Table~\ref{tab:bursts_1820}).

\subsubsection{Analysis of the superburst precursor}

\begin{figure}
\includegraphics[clip]{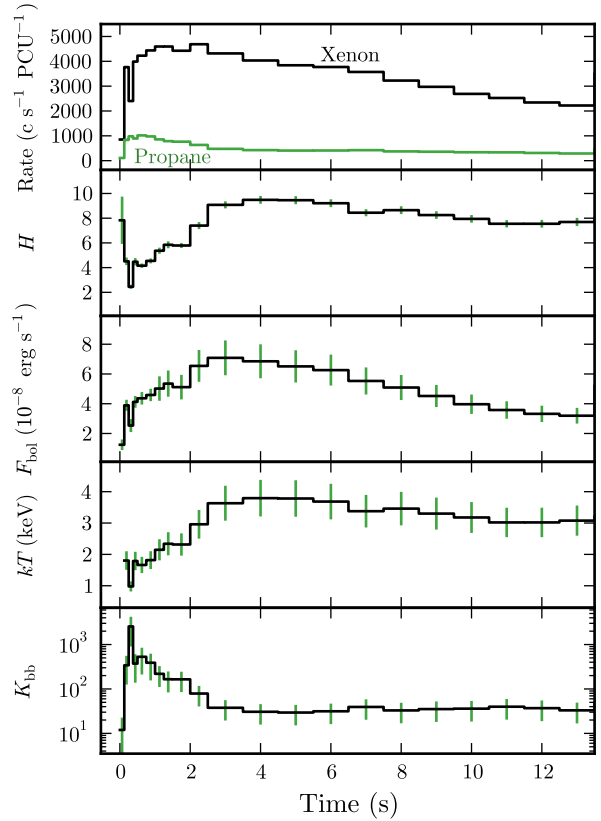}

\caption{\label{fig:precursor1820} Similar to Figure~\ref{fig:PRE-burst}
for the precursor of the 4U~1820--30 superburst.}
\end{figure}
We divide the first $13.5\,\mathrm{s}$ of the superburst into time
intervals, using $0.125\,\mathrm{s}$ bins at the start to resolve
the rise and dip in the count rate at the highest available resolution.
We double the duration of the time intervals several times to get
similar uncertainties in the data points of the entire precursor.
The tail is resolved at $1\,\mathrm{s}$ resolution. We extract $H$
and $I$, and use equations (\ref{eq:hardness_kT-1820}, \ref{eq:ratio_kT-1820})
to derive the black body parameters (Figure~\ref{fig:precursor1820}).
At the start of the precursor we can clearly see $K_{\mathrm{bb}}$
increasing, accompanied by the dip in $kT$. This is indicative of
PRE.

The unabsorbed bolometric black-body flux $F_{\mathrm{bol}}=\frac{R^{2}}{d^{2}}\sigma T^{4}$
can be expressed in terms of $I$ and $kT$ using equations (\ref{eq:hardness_kT-1820},
\ref{eq:ratio_kT-1820}): 
\begin{equation}
F_{\mathrm{bol}}=I\left(\frac{kT}{2\,\mathrm{keV}}\right)^{\Gamma}\,(8.6\pm0.7)\times10^{-12}\,\mathrm{erg\, s^{-1}\, cm^{-2}},\label{eq:Fbol}
\end{equation}
 with $\Gamma=0.2\pm0.2$ for $kT<2\,\mathrm{keV}$ and $\Gamma=0.90\pm0.07$
for $kT\ge2\,\mathrm{keV}$. There is a brief dip in $F_{\mathrm{bol}}$
at the moment of maximum radius expansion (Figure~\ref{fig:precursor1820}).
This is most likely caused by the peak of the Planck spectrum being
outside the observable band when the temperature drops to $\sim1\,\mathrm{keV}$,
and not due to a genuine drop in the bolometric flux.

The light curve as shown in Figure~\ref{fig:precursor1820}, i.e.
the first $13.5\,\mathrm{s}$ of the precursor, has an unabsorbed
bolometric fluence of $(6.8\pm1.1)\times10^{-7}\,\mathrm{erg\, cm^{-2}}$.
After $13.5\,\mathrm{s}$, the light curve rises to the superburst
peak. If we were to consider the precursor as an isolated burst, without
the interruption by the superburst, the flux would continue to decay.
We fit $F_{\mathrm{bol}}$ with an exponential, starting at $t=5.5\,\mathrm{s}$,
when the flux drops below $90\%$ of the peak flux (cf. \citealt{Galloway2008catalog}).
We find an exponential decay time scale $\tau=10\pm3\,\mathrm{s}$.
Using this time scale to extrapolate beyond $13.5\,\mathrm{s}$, we
estimate an additional contribution to the precursor fluence of $(1.5\pm1.0)\times10^{-7}\,\mathrm{erg\, cm^{-2}}$,
giving a total of $(8\pm2)\times10^{-7}\,\mathrm{erg\, cm^{-2}}$.

\subsubsection{Analysis of the second PRE phase}

\begin{figure}
\includegraphics[clip]{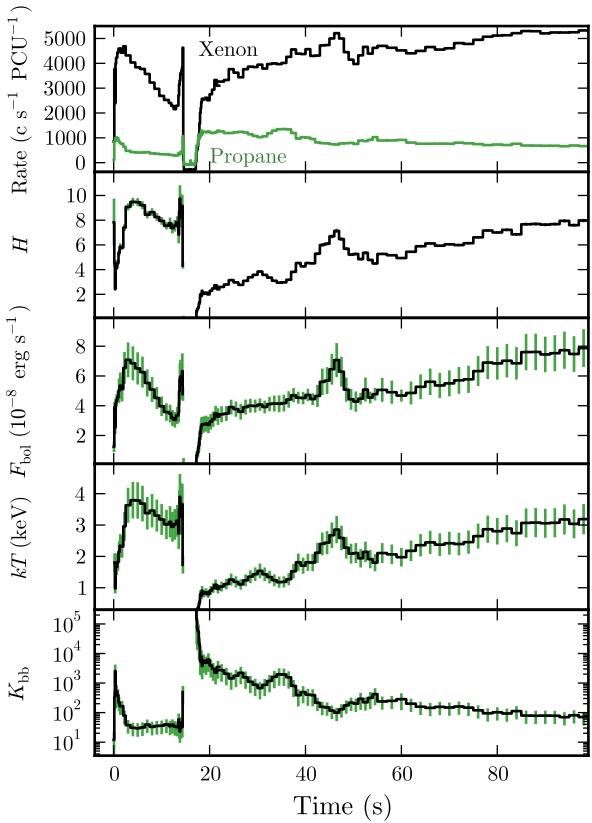}

\caption{\label{fig:sb1820-1} Similar to Figure~\ref{fig:PRE-burst} for
the start of the 4U~1820--30 superburst, including the precursor.
We omit the data points where the xenon rate drops below $0$.}
\end{figure}
The superburst was only spectrally analyzed from $100\,\mathrm{s}$
after the (precursor) onset by \citet{Strohmayer2002}. \citet{Zand2010}
study the first $100\,\mathrm{s}$ at a time resolution of $16\,\mathrm{s}$.
We use our technique to analyze this time interval at higher time
resolution (Figure~\ref{fig:sb1820-1}). Variability is present in
$F_{\mathrm{bol}}$, most notably around $t\simeq47\,\mathrm{s}$,
which exhibits an anticorrelation between $kT$ and $K_{\mathrm{bb}}$.
The variability is present in the signal from all active PCUs.

The superburst shows very strong radius expansion: an order of magnitude
larger than the radius expansion during the precursor. Assuming a
distance of $7.6\pm0.4\,\mathrm{kpc}$ (\citealt{Kuulkers2003}),
we estimate the velocity of the photosphere during expansion by dividing
the change in radius between consecutive data points by the change
in time, and find as largest values $(2.0\pm0.7)\times10^{2}\,\mathrm{km\, s^{-1}}$
during the precursor (at $t=0.3\,\mathrm{s}$ in Figure~\ref{fig:precursor1820})
and $(1.6\pm0.6)\times10^{4}\,\mathrm{km\, s^{-1}}$ during the superexpansion
phase (at $t=14\,\mathrm{s}$ in Figure~\ref{fig:sb1820-1}). Note
that both velocities are each only found from the difference between
two data points. Velocity measurements directly preceding are substantially
lower, indicating strong acceleration. Potentially the velocity of
the photosphere increased further during the superexpansion, when
the signal was lost.

\subsubsection{Comparison of fluence of the precursor and short bursts}

\label{sub:corrections_fluence}\citet{Strohmayer2002} performed
a time-resolved spectral analysis of the superburst from 4U~1820--30
using Standard2 data at $16\,\mathrm{s}$ resolution. They start at
$100\,\mathrm{s}$, because at earlier times the spectral changes
are not well resolved by $16\,\mathrm{s}$ time bins. At $t=100\,\mathrm{s}$
they report $kT\simeq2.6\,\mathrm{keV}$ and $K_{\mathrm{bb}}\simeq115$.
We find at that time $kT=3.2\pm0.5\,\mathrm{keV}$ and $K_{\mathrm{bb}}=(7\pm4)\times10^{1}$,
which differ by $1.1\,\sigma$ and $1.2\,\sigma$, respectively, from
the measurement by \citealt{Strohmayer2002} (see also \citealt{Zand2010}).
The hardness ratio is $H=7.88\pm0.14$, which is at the upper end
of the interval for which the linear correlation between $kT$ and
$H$ describes the data well, whereas at higher values of $H$ the
derived $kT$ values lie somewhat below that line (Figure~\ref{fig:Hardness-kT-1}).

If we take this as a sign that our derived $kT$ values are too large
by a factor $3.2/2.6$, we can scale equation (\ref{eq:hardness_kT-1820})
by the inverse of this factor. Redoing the analysis with reduced temperatures
yields a fluence of $(5.7\pm0.8)\times10^{-7}\,\mathrm{erg\, cm^{-2}}$,
and $(7.8\pm1.5)\times10^{-7}\,\mathrm{erg\, cm^{-2}}$ including
an extrapolated tail. The effect on $K_{\mathrm{bb}}$ can be estimated
from equation (\ref{eq:ratio_kT-1820}) to be approximately a factor
$2$, implying the radius is larger by a factor $1.4$.

The maximum burst fluence we measure is $(4.1\pm0.2)\times10^{-7}\,\mathrm{erg\, cm^{-2}}$
(Table~\ref{tab:bursts_1820}). This is $2.5\,\sigma$ away from
our original fluence measurement and $2.0\,\sigma$ from the one with
the scaled $kT$; here we did not include the estimate of the extrapolated
tail fluence, as this increases the uncertainty. Whereas the inherent
uncertainty in the method we employ does not allow for a strongly
significant difference in the precursor fluence and the maximum burst
fluence, a direct comparison of the integrated burst counts clearly
shows the precursor to be more energetic (Figures \ref{fig:PRE-burst},
\ref{fig:precursor1820}). The normal bursts have at most $(3.37\pm0.07)\times10^{4}\,\mathrm{c\, cm^{-2}}$
(Table~\ref{tab:bursts_1820}), whereas the precursor (without extrapolated
tail) has $(4.59\pm0.06)\times10^{4}\,\mathrm{c\, cm^{-2}}$, which
is a $17\,\sigma$ difference. The precursor has at least $1.36\pm0.03$
times more counts, and between $1.4\pm0.2$ (scaled, no tail) and
$2.0\pm0.5$ (original, with tail) more fluence. From the difference
in count rates, we conclude that the precursor is significantly and
substantially more energetic than the normal bursts.

\subsection{4U~1636--536}

\begin{figure}
\includegraphics{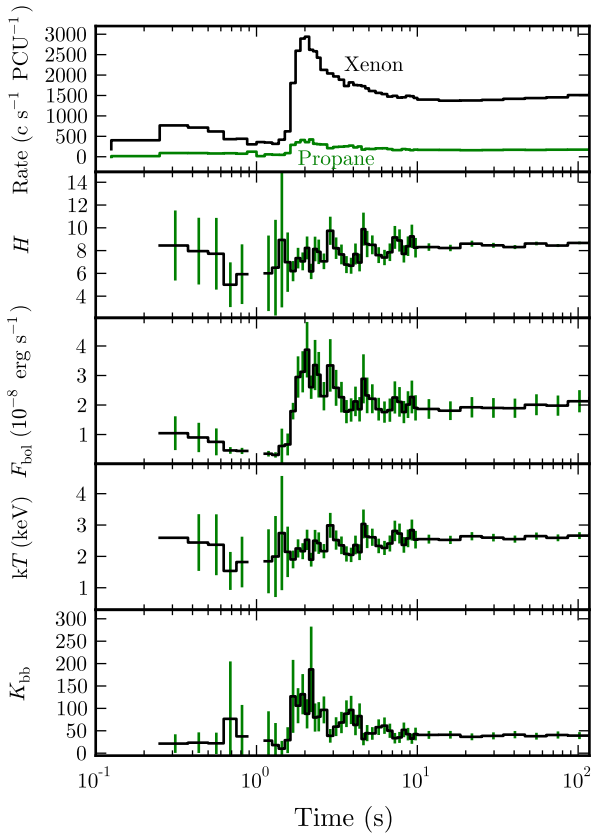}

\caption{\label{fig:sb-1636}Similar to Figure~\ref{fig:PRE-burst} for the
precursor of the 4U~1636--53 superburst. Data points with particularly
large uncertainties in the hardness are omitted.}
\end{figure}
On 2/22/2001 the PCA observed a superburst from 4U~1636--53. The
light curve exhibits a double peak at the onset (Figure~\ref{fig:sb-1636}).
The number of active PCUs with a propane layer was the same as during
the superburst from 4U~1820--30 (3), but the observed propane count
rate was less than half. The uncertainty in the hardness ratio, $H$,
is large in the first few seconds. Coincident with the dip in the
rate, a hint of a dip in $H$ may be present around $1\,\mathrm{s}$
after the onset, indicating a dip in the black-body temperature, which
would be consistent with PRE. The uncertainties, however, are far
too large to draw any conclusions. We applied the same analysis as
for 4U~1820--30, using three PRE bursts detected in the preceding
month with the PCA (Table~\ref{tab:bursts_1636}).
\begin{table}
\caption{\label{tab:bursts_1636}3 PRE Bursts from 4U~1636--53 and a Superburst
Precursor (Bottom)}
\begin{tabular}{r@{\extracolsep{0pt}.}lr@{\extracolsep{0pt}.}l>{\centering}p{1.5cm}>{\centering}p{2.2cm}}
\hline 
Obs& ID & \multicolumn{2}{c}{Time} & Fluence  & Xenon counts \tabularnewline
\multicolumn{2}{c}{} & \multicolumn{2}{c}{(MJD)} & ($10^{-7}$ erg cm$^{-2}$) & ($10^{4}$ c cm$^{-2}$ PCU$^{-1}$)\tabularnewline
\hline 
\multicolumn{2}{c}{50030-02-04-00} & 51937&11615 & $4.93\pm0.12$ & $4.25\pm0.09$\tabularnewline
\multicolumn{2}{c}{50030-02-05-00} & 51942&10027 & $4.78\pm0.14$ & $4.49\pm0.10$\tabularnewline
\multicolumn{2}{c}{50030-02-05-01} & 51941&87561 & $4.39\pm0.10$ & $3.87\pm0.08$\tabularnewline
\hline 
\multicolumn{2}{c}{50030-02-08-01} & 51962&70299 & $3.5\pm0.9$ & $2.69\pm0.06$\tabularnewline
\hline 
\end{tabular}
\end{table}
 We find relations analogous to Equations (\ref{eq:hardness_kT-1820},
\ref{eq:ratio_kT-1820}):

\begin{equation}
kT=(0.307\pm0.012)H\,\mathrm{keV},\label{eq:hardness_kT-1636}
\end{equation}
which requires a systematic uncertainty of $7\%$ to obtain $\chi_{\nu}^{2}=1$,
and 
\begin{equation}
\frac{K_{\mathrm{bb}}}{I}=(4.9\pm0.7)\times10^{-2}\left(\frac{kT}{2\,\mathrm{keV}}\right)^{-\Gamma},\label{eq:ratio_kT-1636}
\end{equation}
with $\Gamma=4.2\pm0.3$ for $kT<2\,\mathrm{keV}$ and $\Gamma=2.6\pm0.3$
for $kT\ge2\,\mathrm{keV}$. These values are consistent with those
obtained for 4U~1820--30. Using these relations we determine the
black-body parameters for the superburst precursor. There is a hint
of weak radius expansion, but the errors are too large to constrain
whether PRE is present (Figure~\ref{fig:sb-1636}).

Because the rise of the superburst emission in the light curve is
slow compared to the superburst from 4U~1820--30, it is more difficult
to separate the precursor flux from the superburst flux. We determine
the total counts and the fluence up to the local minimum in the flux
at $t=18\,\mathrm{s}$ (Table~\ref{tab:bursts_1636}). For the three
short bursts, the counts and fluence at $t>18\,\mathrm{s}$ contribute
at most 15\% of the total. Furthermore, part of the counts and fluence
at $t<18\,\mathrm{s}$ may be attributed to the superburst. Therefore,
we regard the counts and fluence up to $t=18\,\mathrm{s}$, increased
by 15\%, to be an upper limit to the precursor emission. Because of
the large uncertainty in the flux, the fluence is not significantly
different from the fluence of the short bursts, but the counts are
significantly lower.

\section{Discussion}

We devised a method to perform time-resolved spectral analysis of
superburst precursors using data from the anti-coincidence detector
of the PCA on \emph{RXTE}. For a superburst of 4U~1820--30 we clearly
see PRE, but for 4U~1636--53 the data quality is insufficient to
argue whether radius expansion occurred. In this section we first
discuss the possibility for improvements to our spectral analysis
technique. Then we discuss the implications of our results.

\subsection{Improvements to the analysis method}

\label{sub:method improvements}The spectrum of Type I bursts is typically
well fit by a black body (\citealt{swank1977}). Using the propane
and xenon counts we have precisely enough energy ``channels'' to
fit this model. There are improvements over a pure black body that
we cannot implement. Corrections are proposed in the form of a so-called
color correction factor for the black-body temperature of typically
several tens of percents (e.g., \citealt{Suleimanov2010}). Furthermore,
we obtain the net burst spectrum by subtracting the persistent background
from a pre-burst time interval. Here we assume that the pre-burst
emission is dominated by the accretion disk (e.g., \citealt{2002Kuulkers}),
and that this emission remains unchanged during the burst. When there
is strong radius expansion this may no longer be correct. Indeed,
the count rate drops below the pre-burst level during the long PRE
phase in the superburst rise of 4U~1820--30 (Figure~\ref{fig:sb1820-1}).
\citet{Strohmayer2002} find that in the tail the qualitative results
do not differ greatly between assuming a constant persistent flux
and including the persistent spectrum in the fit.

\citet{Strohmayer2002} include in their spectral model for the superburst
of 4U~1820--30 an emission line and an absorption edge. These features
possibly result from reflection of burst emission off the accretion
disk. Later studies have successfully used spectral models including
a disk reflection component (\citealt{Ballantyne2004,Zand2010}).
We do not include these in the model when fitting the normal bursts.
Even though they reach the Eddington limit, the radius expansion is
smaller for the normal bursts than during the superburst. Furthermore,
during the superburst longer time intervals can be chosen for the
spectral analysis, which allows these spectral features to become
significant. This is not the case during the short bursts.

There is considerable spread in the relation between the hardness
ratio and the temperature (Figure~\ref{fig:Hardness-kT-1}). The
width of the distribution is due to the spread in the data points
of individual bursts and the spread between bursts. We investigate
the possibility of improving on the fitted linear relation. The xenon
detector's effective area is well known, but the area is poorly constrained
for the propane layer. An effective area curve, that has the shape
of a power law with a smooth low-energy cut-off, is presented in the
XTE Technical Appendix F. We employ this curve and a black-body model
in an attempt to reproduce the propane counts for the short bursts.
Whereas at low count rates there is reasonable agreement, especially
at the higher count rates during the burst peaks there are strong
deviations from the observed count rates, with a large spread in the
deviations. The effective area curve is, therefore, unusable to improve
the $H-kT$ relation.

\subsection{Precursor from 4U~1820--30}

The start of the PCA superburst from 4U~1820--30 exhibits two instances
of a peak and dip in the light curve (\citealt{Strohmayer2002}).
The first we refer to as the precursor, whereas the second has been
identified as a period of superexpansion (\citealt{Zand2010}, see
Section~\ref{sub:superexpansion}). 

The precursor clearly exhibits PRE (Figure~\ref{fig:precursor1820}).
The peak of the black-body temperature is reached $4\,\mathrm{s}$
after the onset (the ``touch-down point''). The peak values of the
black-body parameters are similar to those of the short bursts (Figure~\ref{fig:PRE-burst}).
The precursor has, however, a longer decay time and a larger fluence.
It is more energetic by a factor $1.4-2$ than the most energetic
of the short bursts we analyzed (Table~\ref{tab:bursts_1820}).

If a sufficiently large column of carbon ignites, burning is predicted
to proceed as a detonation (\citealt{Weinberg2006sb}). A shock is
generated, and travels to the surface (\citealt{Weinberg2007}). This
produces a brief shock breakout peak in the light curve. The shock
heats the atmosphere, which produces a precursor burst. Furthermore,
it triggers the burning of hydrogen and helium, which adds to the
energetics of the precursor (\citealt{Weinberg2007}). \citet{Keek2011}
showed that most of the heat from the shock is released upon the fallback
of shock-accelerated layers. At that moment any available hydrogen/helium
ignites (\citealt{Keek2012}). The heat released by the shock is expected
to be of similar magnitude as that produced by helium burning in the
atmosphere.

The superburst took place at a level of persistent flux where short
bursts are observed from this source (\citealt{Zdziarski2007}). The
amount of helium present in the neutron star atmosphere at the moment
of the superburst depends on the time since the last occurrence of
a short burst. Unfortunately this burst was not observed, but the
fuel column is at most of similar size as the most energetic short
burst. 

Given that we observe the precursor to have a larger fluence than
the short bursts, the burning of helium is insufficient to power the
precursor. We propose shock heating as the most likely source to complete
the precursor energetics. This provides a strong indication that the
superburst indeed proceeded as a detonation, and confirms the scenario
of a shock-induced precursor.

The two peaks in the $2-60\,\mathrm{keV}$ precursor light curve are
caused by PRE; they are not two peaks intrinsic to the (bolometric)
signal. This disproves the suggestion that the first peak is caused
by shock breakout and the second is powered by a helium flash (\citealt{Weinberg2007}).
A scenario where both these events are separated in time by an interval
that is shorter than the $0.125\,\mathrm{s}$ resolution of the data,
such as the dynamical timescale of $\sim10^{-5}\,\mathrm{s}$, is
consistent with the observations (\citealt{Keek2011}).

\subsection{Superexpansion phase}

\label{sub:superexpansion}The 4U~1820--30 superburst is one of only
few bursts to exhibit superexpansion, where briefly the black-body
radius increases by as much as a factor $100$ (\citealt{Zand2010}).
It is speculated to be caused by the ejection of a shell of material
from the neutron star surface. The superexpansion disappears from
the observations when the shell turns optically thin, and a lower
lying layer becomes visible (\citealt{Zand2010}). This lower layer
is moderately expanded and is observed to settle over time. This superburst
has the longest period of moderate expansion observed to date of close
to $1400\,\mathrm{s}$. 

Using our method we can study the superexpansion of the superburst
at much higher time resolution (Figure~\ref{fig:sb1820-1}). Using
a distance of $7.6\,\mathrm{kpc}$ (\citealt{Kuulkers2003}), we find
black-body radii of up to $280\,\mathrm{km}$, and we measure the
rate at which the black-body radius increases to be $(1.6\pm0.6)\times10^{4}\,\mathrm{km\, s^{-1}}$,
which are similar to the values found for two other superexpansion
bursts (\citealt{Zand2010}). Including corrections can lead to $40\%$
higher values (Section~\ref{sub:corrections_fluence}). The rate
of black-body radius increase may trace the velocity of the expanding
photosphere, but deviation of the spectrum from a black body and the
changing optical depth have to be taken into account to determine
the true photospheric velocity. Furthermore, during the strong radius
expansion phase variations in the persistent flux may be present,
which have not been accounted for.

\subsection{Variability in the superburst light curve}

Achromatic variability is present in the tail of the superburst of
4U~1820--30 between $3365-7200\,\mathrm{s}$ after the onset (\citealt{Strohmayer2002}).
This variability is suggested to result from alternate blocking and
reflecting of burst emission by density inhomogeneities in the accretion
disk, that are induced in the superexpansion phase (\citealt{Zand2011}). 

Shortly after the superexpansion period (around $47\,\mathrm{s}$
after onset or $33\,\mathrm{s}$ after the onset of superexpansion),
another peak is present in the flux. The peak is also visible in the
hardness ratio, so this is not achromatic variability and must be
of a different origin. We also find the variability in the black-body
parameters. These were, however, derived assuming a black-body spectrum,
whereas the variability need not be of this nature. For example, the
persistent background may vary, whereas we assume a constant background.
\citet{Zand2011} estimate different timescales for processes in the
disk. The timescale closest to that of the observed variability is
the viscous timescale of approximately $60-950\,\mathrm{s}$ on which
the accretion disk settles.

\subsection{Precursor from 4U~1636--53}

The count rate for the superburst precursor of 4U~1636--53 exhibits
a dip similar to the light curve of the 4U~1820--30 superburst. Only
a weak hint of PRE, however, may be visible in the hardness ratio.
The quality of the data at the onset of the superburst is insufficient
to argue for the presence of absence of radius expansion. The total
photon counts are significantly lower for the precursor than for three
short bursts in the preceeding month. Unlike the precursor of the
4U~1820--30 superburst, therefore, this precursor is not more energetic
than what can be provided by nuclear burning in the atmosphere. \citet{Cumming2006}
derive for this superburst an ignition depth that is just above the
minimum required to generate a shock (\citealt{Keek2011}; see also
\citealt{Weinberg2006sb}), and the models of \citet{Keek2011} that
best fit the light curve do not produce a shock at all. The precursor
would, therefore, mostly be powered by the burning of the available
hydrogen and helium in the envelope, the amount of which depends on
the time since the previous short burst, which is unknown. This may
explain the lack of a powerful precursor and a strong PRE phase.

\subsection{Outlook}

The time resolution of $0.125\,\mathrm{s}$ was insufficient for fully
resolving the start of the radius-expansion phase of the precursor
from 4U~1820--30, and the collecting area was too small to study
the 4U~1636--53 precursor. A future mission such as the Large Observatory
for X-ray Timing (LOFT, \citealt{Feroci2011_LOFT}) is required to
study precursors in detail. This will allow us to confirm the PRE
nature in other superbursts, and provide important information on
the detonation process deep in the neutron star envelope.

\section{Conclusions}

Using data from the propane detector of the PCA on \emph{RXTE}, we
perform the first detailed spectral analysis of superburst precursors.
For the PCA superburst from 4U~1636--53 the data are of insufficient
quality, but for the PCA superburst from 4U~1820--30 the precursor
clearly exhibits PRE. This confirms that the precursor is a single
peak in the (bolometric) emission, and the dip in the peak is due
to the limited band pass of the instrument.

The precursor from 4U~1820--30 is $1.4-2$ times more energetic than
the most powerful short X-ray burst we analyzed from the same source.
This shows that the thermonuclear burning of helium in the atmosphere
is insufficient to power the precursor. We suggest shock heating as
the most likely additional contributor to the energetics. This is
strong support for recent numerical models that predict superbursts
to proceed as a detonation, and that the generated shock deposits
enough heat in the atmosphere to power a bright precursor burst.

With the same technique we study the superexpansion phase of the 4U~1820--30
superburst in greater detail then was previously possible. We find
similar expansion factors as reported for the few other superexpansion
bursts, and we derive expansion velocities of up to $5\%$ of the
speed of light.

\acknowledgements{The author thanks E.\,F.~Brown and J.\,J.\,M.~in~'t~Zand for
comments on this paper, and E.~Kuulkers for helpful discussions.
This paper uses preliminary analysis results from the Multi-INstrument
Burst ARchive (MINBAR), which is supported under the Australian Academy
of Science's Scientific Visits to Europe program, and the Australian
Research Council's Discovery Projects and Future Fellowship funding
schemes. The author thanks the International Space Science Institute
in Bern for hosting an International Team on Type I X-ray bursts.
The author is supported by the Joint Institute for Nuclear Astrophysics
(JINA; grant PHY08-22648), a National Science Foundation Physics Frontier
Center.}

\bibliographystyle{apj}
\bibliography{apj-jour,precursor}

\begin{thebibliography}{44}
\expandafter\ifx\csname natexlab\endcsname\relax\def\natexlab#1{#1}\fi

\bibitem[{{Asai} {et~al.}(2000){Asai}, {Dotani}, {Nagase}, \&
  {Mitsuda}}]{Asai2000}
{Asai}, K., {Dotani}, T., {Nagase}, F., \& {Mitsuda}, K. 2000, \apjs, 131, 571

\bibitem[{{Ballantyne} \& {Strohmayer}(2004)}]{Ballantyne2004}
{Ballantyne}, D.~R., \& {Strohmayer}, T.~E. 2004, \apjl, 602, L105

\bibitem[{{Chou} \& {Grindlay}(2001)}]{cho1}
{Chou}, Y., \& {Grindlay}, J.~E. 2001, \apj, 563, 934

\bibitem[{{Cornelisse} {et~al.}(2000){Cornelisse}, {Heise}, {Kuulkers},
  {Verbunt}, \& {in~'t~Zand}}]{Cornelisse2000}
{Cornelisse}, R., {Heise}, J., {Kuulkers}, E., {Verbunt}, F., \& {in~'t~Zand},
  J.~J.~M. 2000, \aap, 357, L21

\bibitem[{{Cornelisse} {et~al.}(2003){Cornelisse}, {in~'t~Zand}, {Verbunt},
  {Kuulkers}, {Heise}, {den Hartog}, {Cocchi}, {Natalucci}, {Bazzano}, \&
  {Ubertini}}]{Cornelisse2003}
{Cornelisse}, R., {et~al.} 2003, \aap, 405, 1033

\bibitem[{{Cumming}(2003)}]{2003Cumming}
{Cumming}, A. 2003, \apj, 595, 1077

\bibitem[{{Cumming} \& {Bildsten}(2001)}]{Cumming2001}
{Cumming}, A., \& {Bildsten}, L. 2001, \apjl, 559, L127

\bibitem[{{Cumming} {et~al.}(2006){Cumming}, {Macbeth}, {in~'t~Zand}, \&
  {Page}}]{Cumming2006}
{Cumming}, A., {Macbeth}, J., {in~'t~Zand}, J.~J.~M., \& {Page}, D. 2006, \apj,
  646, 429

\bibitem[{{Feroci} {et~al.}(2011){Feroci}, {Stella}, {van der Klis},
  {Courvoisier}, {Hernanz}, {Hudec}, {Santangelo}, {Walton}, {Zdziarski},
  {Barret}, {Belloni}, {Braga}, {Brandt}, {Budtz-J{\o}rgensen}, {Campana}, {den
  Herder}, {Huovelin}, {Israel}, {Pohl}, {Ray}, {Vacchi}, {Zane}, {Argan},
  {Attin{\`a}}, {Bertuccio}, {Bozzo}, {Campana}, {Chakrabarty}, {Costa}, {de
  Rosa}, {Del Monte}, {di Cosimo}, {Donnarumma}, {Evangelista}, {Haas},
  {Jonker}, {Korpela}, {Labanti}, {Malcovati}, {Mignani}, {Muleri},
  {Rapisarda}, {Rashevsky}, {Rea}, {Rubini}, {Tenzer}, {Wilson-Hodge},
  {Winter}, {Wood}, {Zampa}, {Zampa}, {Abramowicz}, {Alpar}, {Altamirano},
  {Alvarez}, {Amati}, {Amoros}, {Antonelli}, {Artigue}, {Azzarello},
  {Bachetti}, {Baldazzi}, {Barbera}, {Barbieri}, {Basa}, {Baykal}, {Belmont},
  {Boirin}, {Bonvicini}, {Burderi}, {Bursa}, {Cabanac}, {Cackett}, {Caliandro},
  {Casella}, {Chaty}, {Chenevez}, {Coe}, {Collura}, {Corongiu}, {Covino},
  {Cusumano}, {D'Amico}, {Dall'Osso}, {de Martino}, {de Paris}, {di Persio},
  {di Salvo}, {Done}, {Dov{\v c}iak}, {Drago}, {Ertan}, {Fabiani}, {Falanga},
  {Fender}, {Ferrando}, {Della Monica Ferreira}, {Fraser}, {Frontera},
  {Fuschino}, {Galvez}, {Gandhi}, {Giommi}, {Godet}, {G{\"o}{\v g}{\"u}{\c s}},
  {Goldwurm}, {G{\"o}tz}, {Grassi}, {Guttridge}, {Hakala}, {Henri}, {Hermsen},
  {Horak}, {Hornstrup}, {in't Zand}, {Isern}, {Kalemci}, {Kanbach}, {Karas},
  {Kataria}, {Kennedy}, {Klochkov}, {Klu{\'z}niak}, {Kokkotas}, {Kreykenbohm},
  {Krolik}, {Kuiper}, {Kuvvetli}, {Kylafis}, {Lattimer}, {Lazzarotto}, {Leahy},
  {Lebrun}, {Lin}, {Lund}, {Maccarone}, {Malzac}, {Marisaldi}, {Martindale},
  {Mastropietro}, {McClintock}, {McHardy}, {Mendez}, {Mereghetti}, {Miller},
  {Mineo}, {Morelli}, {Morsink}, {Motch}, {Motta}, {Mu{\~n}oz-Darias},
  {Naletto}, {Neustroev}, {Nevalainen}, {Olive}, {Orio}, {Orlandini},
  {Orleanski}, {Ozel}, {Pacciani}, {Paltani}, {Papadakis}, {Papitto},
  {Patruno}, {Pellizzoni}, {Petr{\'a}{\v c}ek}, {Petri}, {Petrucci}, {Phlips},
  {Picolli}, {Possenti}, {Psaltis}, {Rambaud}, {Reig}, {Remillard},
  {Rodriguez}, {Romano}, {Romanova}, {Schanz}, {Schmid}, {Segreto}, {Shearer},
  {Smith}, {Smith}, {Soffitta}, {Stergioulas}, {Stolarski}, {Stuchlik},
  {Tiengo}, {Torres}, {T{\"o}r{\"o}k}, {Turolla}, {Uttley}, {Vaughan},
  {Vercellone}, {Waters}, {Watts}, {Wawrzaszek}, {Webb}, {Wilms}, {Zampieri},
  {Zezas}, \& {Ziolkowski}}]{Feroci2011_LOFT}
{Feroci}, M., {et~al.} 2011, Experimental Astronomy, 100

\bibitem[{{Galloway} {et~al.}(2008){Galloway}, {Muno}, {Hartman}, {Psaltis}, \&
  {Chakrabarty}}]{Galloway2008catalog}
{Galloway}, D.~K., {Muno}, M.~P., {Hartman}, J.~M., {Psaltis}, D., \&
  {Chakrabarty}, D. 2008, \apjs, 179, 360

\bibitem[{{Hoffman} {et~al.}(1977){Hoffman}, {Lewin}, \& {Doty}}]{Hoffman1977}
{Hoffman}, J.~A., {Lewin}, W.~H.~G., \& {Doty}, J. 1977, \apjl, 217, L23

\bibitem[{{in~'t~Zand} {et~al.}(2004){in~'t~Zand}, {Cornelisse}, \&
  {Cumming}}]{Zand2004}
{in~'t~Zand}, J.~J.~M., {Cornelisse}, R., \& {Cumming}, A. 2004, \aap, 426, 257

\bibitem[{{in~'t~Zand} {et~al.}(2011{\natexlab{a}}){in~'t~Zand}, {Galloway}, \&
  {Ballantyne}}]{Zand2011}
{in~'t~Zand}, J.~J.~M., {Galloway}, D.~K., \& {Ballantyne}, D.~R.
  2011{\natexlab{a}}, \aap, 525, A111

\bibitem[{{in~'t~Zand} {et~al.}(2003){in~'t~Zand}, {Kuulkers}, {Verbunt},
  {Heise}, \& {Cornelisse}}]{Zand2003}
{in~'t~Zand}, J.~J.~M., {Kuulkers}, E., {Verbunt}, F., {Heise}, J., \&
  {Cornelisse}, R. 2003, \aap, 411, L487

\bibitem[{{in~'t~Zand} {et~al.}(2011{\natexlab{b}}){in~'t~Zand}, {Serino},
  {Kawai}, \& {Heinke}}]{Zand2011ATel}
{in~'t~Zand}, J.~J.~M., {Serino}, M., {Kawai}, N., \& {Heinke}, C.
  2011{\natexlab{b}}, The Astronomer's Telegram, 3625, 1

\bibitem[{{in't Zand} \& {Weinberg}(2010)}]{Zand2010}
{in't Zand}, J.~J.~M., \& {Weinberg}, N.~N. 2010, \aap, 520, A81

\bibitem[{{Jahoda} {et~al.}(2006){Jahoda}, {Markwardt}, {Radeva}, {Rots},
  {Stark}, {Swank}, {Strohmayer}, \& {Zhang}}]{Jahoda2006}
{Jahoda}, K., {Markwardt}, C.~B., {Radeva}, Y., {Rots}, A.~H., {Stark}, M.~J.,
  {Swank}, J.~H., {Strohmayer}, T.~E., \& {Zhang}, W. 2006, \apjs, 163, 401

\bibitem[{{Keek} {et~al.}(2010){Keek}, {Galloway}, {in 't Zand}, \&
  {Heger}}]{Keek2010}
{Keek}, L., {Galloway}, D.~K., {in 't Zand}, J.~J.~M., \& {Heger}, A. 2010,
  \apj, 718, 292

\bibitem[{{Keek} \& {Heger}(2011)}]{Keek2011}
{Keek}, L., \& {Heger}, A. 2011, \apj, 743, 189

\bibitem[{{Keek} {et~al.}(2012){Keek}, {Heger}, \& {in 't Zand}}]{Keek2012}
{Keek}, L., {Heger}, A., \& {in 't Zand}, J.~J.~M. 2012, \apj, 752, 150

\bibitem[{{Keek} \& {in~'t~Zand}(2008)}]{Keek2008int..work}
{Keek}, L., \& {in~'t~Zand}, J.~J.~M. 2008, in Proceedings of the 7th INTEGRAL
  Workshop. 8 - 11 September 2008 Copenhagen, Denmark. Online at
  http://pos.sissa.it/cgi-bin/reader/conf.cgi?confid=67, p.32

\bibitem[{{Kuulkers}(2009)}]{Kuulkers2009ATel}
{Kuulkers}, E. 2009, The Astronomer's Telegram, 2140, 1

\bibitem[{{Kuulkers} {et~al.}(2003){Kuulkers}, {den Hartog}, {in~'t~Zand},
  {Verbunt}, {Harris}, \& {Cocchi}}]{Kuulkers2003}
{Kuulkers}, E., {den Hartog}, P.~R., {in~'t~Zand}, J.~J.~M., {Verbunt},
  F.~W.~M., {Harris}, W.~E., \& {Cocchi}, M. 2003, \aap, 399, 663

\bibitem[{{Kuulkers} {et~al.}(2002{\natexlab{a}}){Kuulkers}, {Homan}, {van der
  Klis}, {Lewin}, \& {M{\'e}ndez}}]{2002Kuulkers}
{Kuulkers}, E., {Homan}, J., {van der Klis}, M., {Lewin}, W.~H.~G., \&
  {M{\'e}ndez}, M. 2002{\natexlab{a}}, \aap, 382, 947

\bibitem[{{Kuulkers} {et~al.}(2004){Kuulkers}, {in~'t~Zand}, {Homan}, {van
  Straaten}, {Altamirano}, \& {van der Klis}}]{2004Kuulkers}
{Kuulkers}, E., {in~'t~Zand}, J., {Homan}, J., {van Straaten}, S.,
  {Altamirano}, D., \& {van der Klis}, M. 2004, in AIP Conf. Proc. 714: X-ray
  Timing 2003: Rossi and Beyond, 257--260

\bibitem[{{Kuulkers} {et~al.}(2002{\natexlab{b}}){Kuulkers}, {in~'t~Zand}, {van
  Kerkwijk}, {Cornelisse}, {Smith}, {Heise}, {Bazzano}, {Cocchi}, {Natalucci},
  \& {Ubertini}}]{Kuulkers2002ks1731}
{Kuulkers}, E., {et~al.} 2002{\natexlab{b}}, \aap, 382, 503

\bibitem[{{Levine} {et~al.}(1996){Levine}, {Bradt}, {Cui}, {Jernigan},
  {Morgan}, {Remillard}, {Shirey}, \& {Smith}}]{Levine1996}
{Levine}, A.~M., {Bradt}, H., {Cui}, W., {Jernigan}, J.~G., {Morgan}, E.~H.,
  {Remillard}, R., {Shirey}, R.~E., \& {Smith}, D.~A. 1996, \apjl, 469, L33+

\bibitem[{{Lewin} {et~al.}(1984){Lewin}, {Vacca}, \& {Basinska}}]{Lewin1984}
{Lewin}, W.~H.~G., {Vacca}, W.~D., \& {Basinska}, E.~M. 1984, \apjl, 277, L57

\bibitem[{{Lewin} {et~al.}(1993){Lewin}, {van Paradijs}, \& {Taam}}]{Lewin1993}
{Lewin}, W.~H.~G., {van Paradijs}, J., \& {Taam}, R.~E. 1993, Space Science
  Reviews, 62, 223

\bibitem[{{Morrison} \& {McCammon}(1983)}]{1983Morrison}
{Morrison}, R., \& {McCammon}, D. 1983, \apj, 270, 119

\bibitem[{{Priedhorsky} \& {Terrell}(1984)}]{Priedhorsky1984}
{Priedhorsky}, W., \& {Terrell}, J. 1984, \apjl, 284, L17

\bibitem[{{Rappaport} {et~al.}(1987){Rappaport}, {Ma}, {Joss}, \&
  {Nelson}}]{Rappaport1987}
{Rappaport}, S., {Ma}, C.~P., {Joss}, P.~C., \& {Nelson}, L.~A. 1987, \apj,
  322, 842

\bibitem[{{Revnivtsev} {et~al.}(2001){Revnivtsev}, {Churazov}, {Gilfanov}, \&
  {Sunyaev}}]{Revnivtsev2001}
{Revnivtsev}, M., {Churazov}, E., {Gilfanov}, M., \& {Sunyaev}, R. 2001, \aap,
  372, 138

\bibitem[{{Stella} {et~al.}(1987){Stella}, {Priedhorsky}, \&
  {White}}]{1987Stella}
{Stella}, L., {Priedhorsky}, W., \& {White}, N.~E. 1987, \apjl, 312, L17

\bibitem[{{Strohmayer} \& {Brown}(2002)}]{Strohmayer2002}
{Strohmayer}, T.~E., \& {Brown}, E.~F. 2002, \apj, 566, 1045

\bibitem[{{Strohmayer} \& {Markwardt}(2002)}]{Strohmayer2002a}
{Strohmayer}, T.~E., \& {Markwardt}, C.~B. 2002, \apj, 577, 337

\bibitem[{{Suleimanov} {et~al.}(2011){Suleimanov}, {Poutanen}, \&
  {Werner}}]{Suleimanov2010}
{Suleimanov}, V., {Poutanen}, J., \& {Werner}, K. 2011, \aap, 527, A139+

\bibitem[{{Swank} {et~al.}(1977){Swank}, {Becker}, {Boldt}, {Holt}, {Pravdo},
  \& {Serlemitsos}}]{swank1977}
{Swank}, J.~H., {Becker}, R.~H., {Boldt}, E.~A., {Holt}, S.~S., {Pravdo},
  S.~H., \& {Serlemitsos}, P.~J. 1977, \apjl, 212, L73

\bibitem[{{Tawara} {et~al.}(1984){Tawara}, {Kii}, {Hayakawa}, {Kunieda},
  {Masai}, {Nagase}, {Inoue}, {Koyama}, {Makino}, {Makishima}, {Matsuoka},
  {Murakami}, {Oda}, {Ogawara}, {Ohashi}, {Shibazaki}, {Tanaka}, {Miyamoto},
  {Tsunemi}, {Yamashita}, \& {Kondo}}]{Tawara1984}
{Tawara}, Y., {et~al.} 1984, \apjl, 276, L41

\bibitem[{{van Paradijs} {et~al.}(1986){van Paradijs}, {Sztajno}, {Lewin},
  {Trumper}, {Vacca}, \& {van der Klis}}]{Paradijs1986}
{van Paradijs}, J., {Sztajno}, M., {Lewin}, W.~H.~G., {Trumper}, J., {Vacca},
  W.~D., \& {van der Klis}, M. 1986, \mnras, 221, 617

\bibitem[{{Weinberg} \& {Bildsten}(2007)}]{Weinberg2007}
{Weinberg}, N.~N., \& {Bildsten}, L. 2007, \apj, 670, 1291

\bibitem[{{Weinberg} {et~al.}(2006){Weinberg}, {Bildsten}, \&
  {Brown}}]{Weinberg2006sb}
{Weinberg}, N.~N., {Bildsten}, L., \& {Brown}, E.~F. 2006, \apjl, 650, L119

\bibitem[{{Wijnands}(2001)}]{Wijnands2001sb}
{Wijnands}, R. 2001, \apjl, 554, L59

\bibitem[{{Zdziarski} {et~al.}(2007){Zdziarski}, {Gierli{\'n}ski}, {Wen}, \&
  {Kostrzewa}}]{Zdziarski2007}
{Zdziarski}, A.~A., {Gierli{\'n}ski}, M., {Wen}, L., \& {Kostrzewa}, Z. 2007,
  \mnras, 377, 1017

\end{thebibliography}

\end{document}